\documentclass[prl,aps,showpacs,epsf,twocolumn]{revtex4}

\def\be{\begin{equation}}
\def\ee{\end{equation}}

\usepackage{graphicx}
\usepackage{amsmath}
\usepackage{bm}
\usepackage{hyperref}

\begin{document}

\title{Dynamics and Thermodynamics of the Low-Temperature\\Strongly Interacting Bose Gas}

\author{Nir Navon$^{1}$\footnote{Electronic address: navon@ens.fr}, Swann Piatecki$^{2}$\footnote{Electronic address: swann.piatecki@ens.fr}, Kenneth G\"{u}nter$^1$, Benno Rem$^1$, Trong Canh Nguyen$^1$, Fr\'ed\'eric Chevy$^1$, Werner Krauth$^2$ and Christophe Salomon$^1$}

\affiliation{$^1$Laboratoire Kastler Brossel, CNRS, UPMC,
\'Ecole Normale Sup\'erieure, 24 rue Lhomond, 75005 Paris, France\\
$^2$Laboratoire de Physique Statistique, \'Ecole Normale Sup\'erieure, UPMC, Univ. Paris Diderot, CNRS, 24 rue Lhomond, 75005 Paris, France}

\date{\today}

\begin{abstract}
We measure the zero-temperature equation of state of a homogeneous Bose gas of $^7$Li atoms by analyzing the \emph{in-situ} density distributions of trapped samples. For increasing repulsive interactions our data shows a clear departure from mean-field theory and provides a quantitative test of the many-body corrections first predicted in 1957 by Lee, Huang and Yang. We further probe the dynamic response of the Bose gas to a varying interaction strength and compare it to simple theoretical models. We deduce a lower bound for the value of the universal constant $\xi>0.44(8)$ that would characterize the universal Bose gas at the unitary limit. 
\end{abstract}

\maketitle

From sand piles to neuronal networks, electrons in metals and quantum liquids, one of the greatest challenges in modern physics is to understand the behavior of strongly interacting systems.
Recent advances in quantum gas experiments have made it possible to test with unprecedented accuracy complex many-body theories beyond a simple mean-field description. 
Strong correlations in quantum gases can be achieved for example by loading the atoms into an optical lattice, where the enhanced role of the interaction with respect to the kinetic energy can drive the gas into a strongly correlated Mott insulating state \cite{greiner2002quantum}. Another possibility is to reduce the dimensionality to increase quantum fluctuations in the system \cite{bloch2008many}. In the fundamental case of a homogeneous three-dimensional Bose gas, a famous example is superfluid $^4$He, where the short-range interaction potential plays a crucial role. For dilute ultracold atomic vapors, the properties of the strongly interacting Bose gas remain largely to be explored.
Magnetic Feshbach resonances make it possible to tune the atom-atom interactions, characterized by the $s$-wave scattering length $a$ \cite{inouye1998observation,roberts2000magnetic}. However, first studies of the strongly interacting regime were plagued by severe inelastic atom loss for Bose systems \cite{fedichev1996three}.
In the last two years, experiments at JILA and Rice have revived interest in these systems \cite{papp2008bragg,pollack2009extreme}. 
Here we report on a quantitative measurement of the thermodynamic Equation of State (EoS) of a strongly interacting atomic Bose gas in the low-temperature limit. 
In a first part, we restrict ourselves to a regime with negligible atom loss ($a/a_0\leq 2150$, $a_0$ being the Bohr radius), and our EoS reveals the first beyond mean-field correction due to quantum fluctuations. We perform Quantum Monte Carlo (QMC) simulations to support our experimental findings. In a second part we discuss the dynamics of a Bose gas brought to the unitary regime. 
\begin{figure}[h!]
\centerline{\includegraphics[width=0.95\columnwidth]{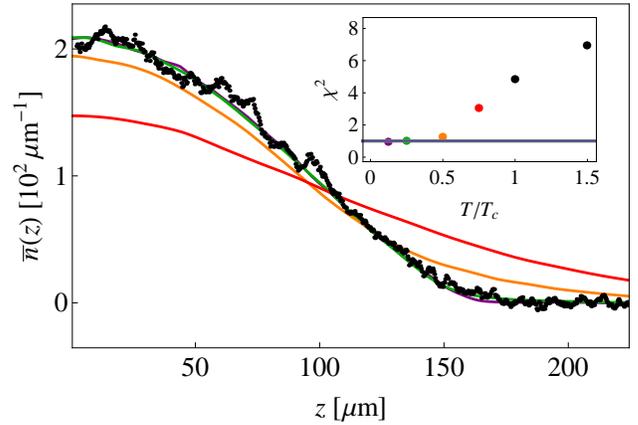}}
\caption{(Color online) Doubly-integrated density profile of a trapped Bose gas at a scattering length $a/a_0=2150$. The average over 5 experimental images is shown in black points. The QMC predictions for $3.9\;10^4$ atoms are plotted in solid line for $T/T_c=0.75$ in red, $0.5$ in orange, $0.25$ in green and $0.125$ in purple. Inset: $\chi^2$ deviation per degree of freedom of a single experimental density profile with QMC results at different temperatures. The excellent agreement between experimental profiles and QMC validates the zero-temperature assumption for the EoS measurement.} \label{plotimagesBEC}
\end{figure}

A seminal result for the thermodynamics of the dilute Bose gas is the expansion of the ground state energy (per volume $V$) in terms of the canonical gas parameter $na^3$, first obtained in the late '50s 
\cite{lee1957eigenvalues}:
\begin{equation}\label{EnergyExpansion}
\frac{E}{V} = \frac{gn^2}{2}\left(1+\frac{128}{15\sqrt{\pi}}\sqrt{na^3} + \ldots \right),
\end{equation}
where $n$ is the density of the gas, and $g=4\pi\hbar^2a/m$ the coupling constant for particles with mass $m$.
The first term in Eq.\ref{EnergyExpansion} is the mean-field energy, while the Lee-Huang-Yang (LHY) correction, proportional to $\sqrt{na^3}$, is due to quantum fluctuations \cite{lee1957eigenvalues}. Up to this order, the expansion is \emph{universal}, in the sense that it solely depends on the gas parameter $na^3$ and not on microscopic details of the interaction potential \cite{brueckner1957bose,beliaev1958application,lieb1963simplified}. The term beyond LHY involves the first \emph{non-universal} correction \cite{braaten2002dilute}. Despite its fundamental importance, this expansion has never been checked experimentally for an atomic Bose gas.

\begin{figure}
\centerline{\includegraphics[width=0.9\columnwidth]{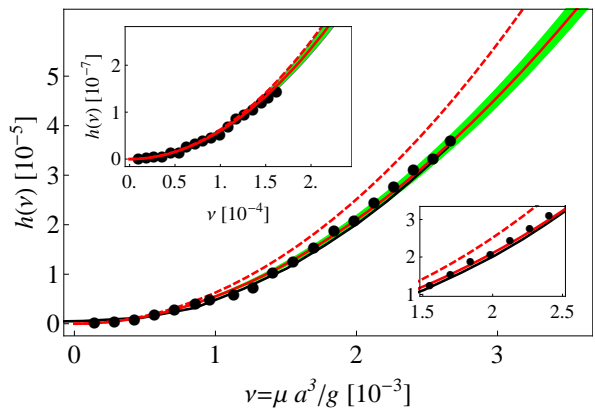}}
\caption{(Color online) Equation of State of the homogeneous Bose gas expressed as the normalized pressure $h$ as a function of the gas parameter $\nu$. The gas samples for the data shown in the main panel (resp. upper inset) have been prepared at scattering lengths of $a/a_0=1450$ and $2150$ (resp. $a/a_0=700$). The red solid line corresponds to the LHY prediction, and the red dashed line, to the mean-field EoS $h(\nu)=2\pi\nu^2$. The solid black line represents the QMC EoS at $T/T_c=0.25$ and is nearly indistinguishable from the LHY EoS (see lower inset). The green area delimits the uncertainty of $5\,\%$ on the value of $a$.} \label{FigEOS}
\end{figure}
Our experimental setup was described in \cite{nascimbene2009pol}. Starting from a $^7$Li cloud in a magneto-optical trap, we optically pump the atoms into the $|F=2,m_F=2\rangle$ hyperfine state and transfer them into a magnetic Ioffe trap. After evaporative cooling to a temperature of $\sim 4\,\mu$K, the atoms are loaded into a hybrid magnetic/optical trap and then transferred to the $|F=1,m_F=1\rangle$ state. The radial optical confinement of the trap is provided by a single laser beam of waist $35\,\mu$m operating at a wavelength 1073\,nm, while the weak axial confinement is enhanced by an additional magnetic curvature. We apply a homogeneous magnetic field to tune the interaction strength by means of a wide Feshbach resonance around 740\,G. The final stage of evaporation in the optical trap is carried out at a bias field of 717\,G, where the scattering length has a value of about $200\,a_0$, and results in a Bose-Einstein condensate of $\sim6\times10^4$ atoms with no discernible thermal part. In the final configuration the trapping frequencies are given by $\omega_r=2\pi\times345(20)$\,Hz in the radial and $\omega_z=2\pi\times 18.5(1)$\,Hz in the axial direction. The magnetic bias field is then adiabatically ramped to the vicinity of the Feshbach resonance and the density distribution is recorded using \emph{in-situ} absorption imaging. 

For the measurement of the EoS, we follow the method of \cite{cheng2007trapped,shin2008determination,ho2009obtaining,nascimb2010exploring,navon2010equation}. Accordingly, the local pressure $P(z)$ along the symmetry axis of a harmonically trapped gas is related to the doubly-integrated \emph{in-situ} density profile $\bar{n}(z) = \int\mathrm{d}x\,\mathrm{d}y\,n(x,y,z)$:
\begin{equation}
\label{Pn}
P(\mu_{z}) = \frac{m\omega_r^2}{2\pi}\bar{n}(z) \,.
\end{equation}
This formula relies on the local-density approximation in which the local chemical potential is defined as $\mu_z = \mu_0-V(0,0,z)$, where $\mu_0$ is the global chemical potential of the gas, and $V({\bf r})$ is the harmonic trapping potential.

As the EoS critically depends on the scattering length, a precise knowledge of the latter close to the Feshbach resonance of the $|1,1\rangle$ state is essential. In view of the discrepancy between two recent works \cite{pollack2009extreme,gross2010nuclear}, we have independently calibrated the scattering length $a(B)$ as a function of magnetic field $B$ by radio-frequency molecule association spectroscopy \cite{zirbel2008heteronuclear}, as described in the supplementary material.
To measure the pressure at different interaction strengths we have selected images with atom numbers in the range of $3$ -- $4\times10^4$ in order to avoid high optical densities during absorption imaging while keeping a good signal-to-noise ratio. A total of 50 images are used, spanning values of $a/a_0$ from $700$ to $2150$. 
We calibrate the relation between the integrated optical density and the pressure of the gas at weak interaction, well described by mean-field theory (inset of 
Fig.\ref{FigEOS}). The density profiles then generate the EoS (\ref{Pn}). 
The global chemical potential $\mu_0$ remains to be determined. For this work, we infer $\mu_0$ self-consistently in a model-independent way from the density profiles (see the supplementary material).

In the dilute limit $na^3\ll1$, where the EoS is universal, the grand-canonical EoS of the homogeneous Bose gas at zero temperature can dimensionally be written as 
\begin{equation}
\label{eqofstate}
P(\mu,a) = \frac{\hbar^2}{ma^5} \cdot h\left(\nu\equiv\frac{\mu}{g}a^3\right),
\end{equation}
where $\nu$ is the (grand-canonical) gas parameter. 
The experimental data points for $h(\nu)$ are displayed in Fig.~\ref{FigEOS}. We observe a clear departure of the EoS from the mean-field prediction (dashed red line). At the largest measured value of $\nu=2.8\times 10^{-3}$ our data show a reduction of $20\,\%$ of the pressure with respect to the mean-field result $h(\nu)=2\pi\nu^2$. 
In Fig.~\ref{FigEOS} we compare the measurements to the mean-field EoS, the LHY result (translated to the grand-canonical ensemble), and our QMC calculation. 
We observe that LHY theory accurately describes our experimental data and is hardly distinguishable from the QMC in the studied range of interaction strength, a point already put forward in a diffusion Monte Carlo simulation at even higher values of the gas parameter \cite{giorgini1999ground}. We can quantify the deviation of our data from mean-field theory by fitting the measured EoS with a function that includes a correction of order $\sqrt{na^3}$. For this purpose we convert the energy $E/N = 2\pi\hbar^2/(ma^2)\cdot na^3\left(1+\alpha (na^3)^{1/2}+...\right)$ to the grand-canonical EoS (see the supplementary material) and use $\alpha$ as a fit parameter in the resulting pressure $P(\mu)$. The fit yields the value $\alpha=4.5(7)$, which is in excellent agreement with the theoretical result $128/(15\sqrt{\pi}) \approx 4.81$ in Eq.~(\ref{EnergyExpansion}).
Together with the measurement with composite bosons of \cite{navon2010equation}, this provides a striking check of the universality predicted by the expansion (\ref{EnergyExpansion}) up to order $\sqrt{na^3}$ \cite{leyronas2007superfluid}. 

In the above interpretation we assumed that the zero-temperature regime has effectively been reached. To check this crucial assumption, we have performed finite-temperature path-integral Quantum Monte Carlo simulations \cite{krauth1996quantum} in the anisotropic harmonic trap geometry of the experiment with continuous space variables. The experimental atom number can be reached without difficulty and pair interactions are described by a pseudo-potential.
All the thermodynamic properties of the gas at finite temperature are obtained to high precision and without systematic errors. From a comparison of the experimental density distributions with the QMC profiles at different temperatures (see Fig.\ref{plotimagesBEC}), we find that the clouds in our experiments are colder than 0.25 $T_c$, where $T_c$ is the condensation temperature of the ideal Bose gas. Thus thermal effects lead to an error in the EoS which is much smaller than the statistical error bars in Fig.\ref{FigEOS}. 

\begin{figure}[h!]
\centerline{\includegraphics[width=0.9\columnwidth]{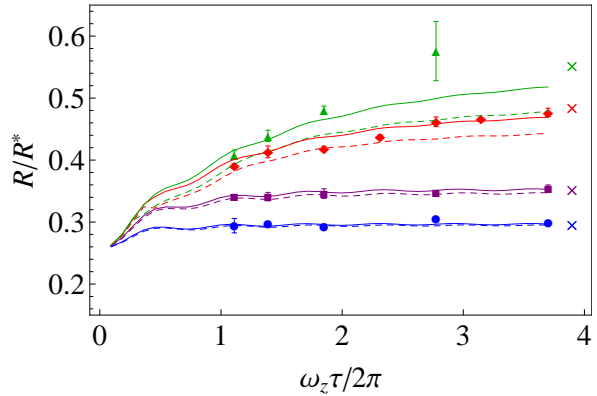}}
\caption{(Color online) Radius $R$ of the Bose gas as a function of the duration $\tau$ of the interaction sweep. The radius $R$ is normalized to the radius $R^*=a_\textrm{ho}(15\lambda^2 N)^{1/5}$ (where $a_\textrm{ho}=\left(\hbar/m\omega_z\right)^{1/2}$ and $\lambda=\omega_r/\omega_z$).
$N$ is the measured atom number at the end of each sweep. The final values of $a/a_0$ are $380$ (blue dots), $840$ (purple squares), $2940$ (red diamonds) and $4580$ (green triangles). The solid (dashed) lines show the solution of a variational hydrodynamic approach (mean-field scaling solutions). The crosses show the predicted equilibrium beyond mean-field radii. \label{rampspeed}}
\end{figure}

We now study the dynamics of a Bose gas subjected to a time-dependent interaction strength towards the strongly interacting regime. Besides its own interest, the investigation of this problem is important to assess the adiabaticity of the interaction sweep in the measurements described above. At strong interactions the enhanced three-body loss rate limits the practical duration of the sweep. In that case, a violation of adiabaticity could lead to non-equilibrium density profiles that distort the measured EoS. To address the question of adiabaticity, we investigate the axial size of the Bose gas after a sweep of the scattering length. In Fig.~\ref{rampspeed} we plot the cloud size determined by a Thomas-Fermi fit as a function of the sweep duration. The magnetic field is ramped approximately linearly in time, sweeping $a/a_0$ from an initial value of $200$ to different final values. Besides the experimental data we present theoretical results from a mean-field scaling solution \cite{castin1996bose,kagan1997evolution} and from a solution of the hydrodynamic equations incorporating the LHY EoS based on a variational scaling ansatz \cite{inprep}. The latter shows a remarkable agreement with our experimental data for $a \le 3000\,a_0$. For scattering lengths $a/a_0 \le 840$ the radius is nearly constant for sweep durations $\tau\omega_z/(2\pi) >1.5$ ($\tau>80$\,ms), indicating that the cloud follows the interaction strength adiabatically. For the largest value employed in the EoS study ($a/a_0=2150$), the atom loss is less than $4$\,\% and the cloud size after the $\tau=150$\,ms sweep ($\tau\omega_z/(2\pi)\gtrsim 2.8$) is 2.5\,\% smaller than the equilibrium value. We have corrected for this systematic effect by rescaling the measured density $n_0$ for the determination of the EoS, $\bar{n}=\eta^{-1}\bar{n}_0(\eta z)$ (with $\eta=0.975$ for $a/a_0=2150$).

\begin{figure}[h!]
\centerline{\includegraphics[width=0.9\columnwidth]{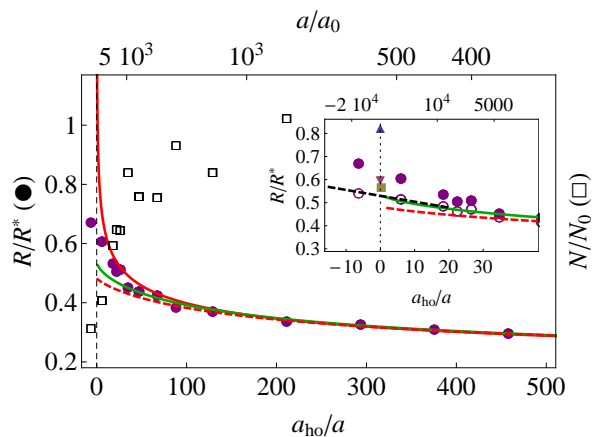}}
\caption{(Color online) Normalized cloud radius $R_\textrm{TF}/R^*$ (filled purple circles) and normalized atom number (open black squares) as a function of the inverse scattering length $a_\textrm{ho}/a$ at the end of a $75$\,ms magnetic field sweep. The static mean-field prediction is plotted in solid red line, the mean-field scaling solution in dashed red, and the beyond mean-field scaling ansatz in solid green line. Inset: Zoom around the unitary limit. Predictions for the universal constant $\xi$ are shown in up triangle \cite{cowell2002cold}, down triangle \cite{song2009ground} and square \cite{lee2010universality}. The filled (empty) circles correspond to the radii normalized to the final (initial) atom number. The dotted black line is the linear interpolation at unitarity.} \label{DynBEC}
\end{figure}

The properties of the Bose gas for large values of $na^3$ constitute a challenging open problem. Due to the experimental limitation imposed by three-body recombination, we access this region with a shorter sweep of duration $\tau\omega_z/(2\pi)=1.35$ ($\tau\approx 75$\,ms). In Fig.~\ref{DynBEC} we plot the normalized radius of the Bose gas as a function of the inverse scattering length $a_\textrm{ho}/a$. 
Deep in the mean-field regime ($a \lesssim 800\,a_0$) the ramp is adiabatic as the data matches the equilibrium Thomas-Fermi prediction. As the scattering length is increased, both non-adiabaticity and beyond mean-field effects become important. A departure from the equilibrium result becomes evident above a scattering length of $\simeq 2000\,a_0$. Taking into account the mean-field dynamics gives already an improved description of our data (red dashed line). Even better agreement (up to values of $a/a_0\simeq 5\,000$) is obtained with the variational approach incorporating the LHY correction as presented above (green solid line) \cite{inprep}.  

Further insight is gained by considering data taken right on the Feshbach resonance and beyond. Because of the low densities of our samples, only half of the atoms are lost at the end of the sweep to the resonance (see squares in Fig.~\ref{DynBEC}). Universal thermodynamics at unitarity (where $a$ diverges) has been conjectured for quantum gases \cite{ho2004universal} and successfully checked experimentally for Fermi gases \cite{inguscio2006ultracold}. In the case of bosonic atoms the existence of a many-body universal state at unitarity is still unknown. If universality proved to be correct, the only relevant length scale would be the interparticle spacing $n^{-1/3}$ and the EoS would take the form $\mu\propto \frac{\hbar^2}{m}n^{2/3}$. Up to a numerical factor, this EoS is identical to that of an ideal Fermi gas and we can write $\mu=\xi E_\mathrm{F}$ (where $E_\mathrm{F}=\hbar^2/2m\:(6\pi^2 n)^{2/3}$). When the scattering length is increased, the cloud is expected to grow in size. Due to the finite response time of the gas, it is reasonable to assume that the measured radius $R$ is smaller than the equilibrium radius. From this inequality we deduce a lower bound for the value of $\xi$ by interpolating our data at unitarity as presented in the inset of Fig.\ref{DynBEC}: $\xi>0.44(8)$ \cite{unitaryradiusscaling}. 
This bound is satisfied for the predictions $\xi=0.66$ \cite{lee2010universality} and for the upper bounds from variational calculations, $0.80$ \cite{song2009ground} and $2.93$ \cite{cowell2002cold}. 

Future work could focus on the measurement of the condensate fraction since the quantum depletion is expected to be as large as $\sim8\,\%$ for our most strongly interacting samples in equilibrium, and on finite temperature thermodynamic properties. Our measurements on resonance as well as future theoretical studies should give crucial insights on the unitary Bose gas.

We are grateful to Y. Castin, F. Werner, X. Leyronas, S. Stringari, S. Giorgini, S. Pilati for discussions and to C. Lancien, I. Lahlou, S. El-Ghazzal, T. Vu and 
L. Bernard for experimental assistance. We thank J. Dalibard and S. Nascimb\`ene for useful comments on the manuscript. We acknowledge support from ESF Euroquam (FerMix), ANR FABIOLA, R\'egion Ile de France (IFRAF), ERC and Institut Universitaire de France.
 
\bibliographystyle{apsrev}

\begin{thebibliography}{32}
\expandafter\ifx\csname natexlab\endcsname\relax\def\natexlab#1{#1}\fi
\expandafter\ifx\csname bibnamefont\endcsname\relax
  \def\bibnamefont#1{#1}\fi
\expandafter\ifx\csname bibfnamefont\endcsname\relax
  \def\bibfnamefont#1{#1}\fi
\expandafter\ifx\csname citenamefont\endcsname\relax
  \def\citenamefont#1{#1}\fi
\expandafter\ifx\csname url\endcsname\relax
  \def\url#1{\texttt{#1}}\fi
\expandafter\ifx\csname urlprefix\endcsname\relax\def\urlprefix{URL }\fi
\providecommand{\bibinfo}[2]{#2}
\providecommand{\eprint}[2][]{\url{#2}}

\bibitem[{\citenamefont{Greiner et~al.}(2002)\citenamefont{Greiner, Mandel,
  Esslinger, Hansch, and Bloch}}]{greiner2002quantum}
\bibinfo{author}{\bibfnamefont{M.}~\bibnamefont{Greiner}},
  \bibinfo{author}{\bibfnamefont{O.}~\bibnamefont{Mandel}},
  \bibinfo{author}{\bibfnamefont{T.}~\bibnamefont{Esslinger}},
  \bibinfo{author}{\bibfnamefont{T.}~\bibnamefont{Hansch}}, \bibnamefont{and}
  \bibinfo{author}{\bibfnamefont{I.}~\bibnamefont{Bloch}},
  \bibinfo{journal}{Nature} \textbf{\bibinfo{volume}{415}}, \bibinfo{pages}{39}
  (\bibinfo{year}{2002}).

\bibitem[{\citenamefont{Bloch et~al.}(2008)\citenamefont{Bloch, Dalibard, and
  Zwerger}}]{bloch2008many}
\bibinfo{author}{\bibfnamefont{I.}~\bibnamefont{Bloch}},
  \bibinfo{author}{\bibfnamefont{J.}~\bibnamefont{Dalibard}}, \bibnamefont{and}
  \bibinfo{author}{\bibfnamefont{W.}~\bibnamefont{Zwerger}},
  \bibinfo{journal}{Rev. Mod. Phys.} \textbf{\bibinfo{volume}{80}},
  \bibinfo{pages}{885} (\bibinfo{year}{2008}).

\bibitem[{\citenamefont{Inouye et~al.}(1998)\citenamefont{Inouye, Andrews,
  Stenger, Miesner, Stamper-Kurn, and Ketterle}}]{inouye1998observation}
\bibinfo{author}{\bibfnamefont{S.}~\bibnamefont{Inouye}},
  \bibinfo{author}{\bibfnamefont{M.}~\bibnamefont{Andrews}},
  \bibinfo{author}{\bibfnamefont{J.}~\bibnamefont{Stenger}},
  \bibinfo{author}{\bibfnamefont{H.}~\bibnamefont{Miesner}},
  \bibinfo{author}{\bibfnamefont{D.}~\bibnamefont{Stamper-Kurn}},
  \bibnamefont{and} \bibinfo{author}{\bibfnamefont{W.}~\bibnamefont{Ketterle}},
  \bibinfo{journal}{Nature} \textbf{\bibinfo{volume}{392}},
  \bibinfo{pages}{151} (\bibinfo{year}{1998}).

\bibitem[{\citenamefont{Roberts et~al.}(2000)\citenamefont{Roberts, Claussen,
  Cornish, and Wieman}}]{roberts2000magnetic}
\bibinfo{author}{\bibfnamefont{J.}~\bibnamefont{Roberts}},
  \bibinfo{author}{\bibfnamefont{N.}~\bibnamefont{Claussen}},
  \bibinfo{author}{\bibfnamefont{S.}~\bibnamefont{Cornish}}, \bibnamefont{and}
  \bibinfo{author}{\bibfnamefont{C.}~\bibnamefont{Wieman}},
  \bibinfo{journal}{Phys. Rev. Lett.} \textbf{\bibinfo{volume}{85}},
  \bibinfo{pages}{728} (\bibinfo{year}{2000}).

\bibitem[{\citenamefont{Fedichev et~al.}(1996)\citenamefont{Fedichev, Reynolds,
  and Shlyapnikov}}]{fedichev1996three}
\bibinfo{author}{\bibfnamefont{P.}~\bibnamefont{Fedichev}},
  \bibinfo{author}{\bibfnamefont{M.}~\bibnamefont{Reynolds}}, \bibnamefont{and}
  \bibinfo{author}{\bibfnamefont{G.}~\bibnamefont{Shlyapnikov}},
  \bibinfo{journal}{Phys. Rev. Lett.} \textbf{\bibinfo{volume}{77}},
  \bibinfo{pages}{2921} (\bibinfo{year}{1996}).

\bibitem[{\citenamefont{Papp et~al.}(2008)\citenamefont{Papp, Pino, Wild,
  Ronen, Wieman, Jin, and Cornell}}]{papp2008bragg}
\bibinfo{author}{\bibfnamefont{S.}~\bibnamefont{Papp}},
  \bibinfo{author}{\bibfnamefont{J.}~\bibnamefont{Pino}},
  \bibinfo{author}{\bibfnamefont{R.}~\bibnamefont{Wild}},
  \bibinfo{author}{\bibfnamefont{S.}~\bibnamefont{Ronen}},
  \bibinfo{author}{\bibfnamefont{C.}~\bibnamefont{Wieman}},
  \bibinfo{author}{\bibfnamefont{D.}~\bibnamefont{Jin}}, \bibnamefont{and}
  \bibinfo{author}{\bibfnamefont{E.}~\bibnamefont{Cornell}},
  \bibinfo{journal}{Phys. Rev. Lett.} \textbf{\bibinfo{volume}{101}},
  \bibinfo{pages}{135301} (\bibinfo{year}{2008}).

\bibitem[{\citenamefont{Pollack et~al.}(2009)\citenamefont{Pollack, Dries,
  Junker, Chen, Corcovilos, and Hulet}}]{pollack2009extreme}
\bibinfo{author}{\bibfnamefont{S.}~\bibnamefont{Pollack}},
  \bibinfo{author}{\bibfnamefont{D.}~\bibnamefont{Dries}},
  \bibinfo{author}{\bibfnamefont{M.}~\bibnamefont{Junker}},
  \bibinfo{author}{\bibfnamefont{Y.}~\bibnamefont{Chen}},
  \bibinfo{author}{\bibfnamefont{T.}~\bibnamefont{Corcovilos}},
  \bibnamefont{and} \bibinfo{author}{\bibfnamefont{R.}~\bibnamefont{Hulet}},
  \bibinfo{journal}{Phys. Rev. Lett.} \textbf{\bibinfo{volume}{102}},
  \bibinfo{pages}{90402} (\bibinfo{year}{2009}).

\bibitem[{\citenamefont{Lee et~al.}(1957)\citenamefont{Lee, Huang, and
  Yang}}]{lee1957eigenvalues}
\bibinfo{author}{\bibfnamefont{T.}~\bibnamefont{Lee}},
  \bibinfo{author}{\bibfnamefont{K.}~\bibnamefont{Huang}}, \bibnamefont{and}
  \bibinfo{author}{\bibfnamefont{C.}~\bibnamefont{Yang}},
  \bibinfo{journal}{Phys. Rev.} \textbf{\bibinfo{volume}{106}},
  \bibinfo{pages}{1135} (\bibinfo{year}{1957}).

\bibitem[{\citenamefont{Brueckner and Sawada}(1957)}]{brueckner1957bose}
\bibinfo{author}{\bibfnamefont{K.}~\bibnamefont{Brueckner}} \bibnamefont{and}
  \bibinfo{author}{\bibfnamefont{K.}~\bibnamefont{Sawada}},
  \bibinfo{journal}{Phys. Rev.} \textbf{\bibinfo{volume}{106}},
  \bibinfo{pages}{1117} (\bibinfo{year}{1957}).

\bibitem[{\citenamefont{Beliaev}(1958)}]{beliaev1958application}
\bibinfo{author}{\bibfnamefont{S.}~\bibnamefont{Beliaev}},
  \bibinfo{journal}{JETP} \textbf{\bibinfo{volume}{7}}, \bibinfo{pages}{289}
  (\bibinfo{year}{1958}).

\bibitem[{\citenamefont{Lieb}(1963)}]{lieb1963simplified}
\bibinfo{author}{\bibfnamefont{E.}~\bibnamefont{Lieb}}, \bibinfo{journal}{Phys.
  Rev.} \textbf{\bibinfo{volume}{130}}, \bibinfo{pages}{2518}
  (\bibinfo{year}{1963}).

\bibitem[{\citenamefont{Braaten et~al.}(2002)\citenamefont{Braaten, Hammer, and
  Mehen}}]{braaten2002dilute}
\bibinfo{author}{\bibfnamefont{E.}~\bibnamefont{Braaten}},
  \bibinfo{author}{\bibfnamefont{H.}~\bibnamefont{Hammer}}, \bibnamefont{and}
  \bibinfo{author}{\bibfnamefont{T.}~\bibnamefont{Mehen}},
  \bibinfo{journal}{Phys. Rev. Lett.} \textbf{\bibinfo{volume}{88}},
  \bibinfo{pages}{40401} (\bibinfo{year}{2002}).

\bibitem[{\citenamefont{Nascimbene et~al.}(2009)\citenamefont{Nascimbene,
  Navon, Jiang, Tarruell, Teichmann, Mckeever, Chevy, and
  Salomon}}]{nascimbene2009pol}
\bibinfo{author}{\bibfnamefont{S.}~\bibnamefont{Nascimbene}},
  \bibinfo{author}{\bibfnamefont{N.}~\bibnamefont{Navon}},
  \bibinfo{author}{\bibfnamefont{K.}~\bibnamefont{Jiang}},
  \bibinfo{author}{\bibfnamefont{L.}~\bibnamefont{Tarruell}},
  \bibinfo{author}{\bibfnamefont{M.}~\bibnamefont{Teichmann}},
  \bibinfo{author}{\bibfnamefont{J.}~\bibnamefont{Mckeever}},
  \bibinfo{author}{\bibfnamefont{F.}~\bibnamefont{Chevy}}, \bibnamefont{and}
  \bibinfo{author}{\bibfnamefont{C.}~\bibnamefont{Salomon}},
  \bibinfo{journal}{Phys. Rev. Lett.} \textbf{\bibinfo{volume}{103}},
  \bibinfo{pages}{170402} (\bibinfo{year}{2009}).

\bibitem[{\citenamefont{Cheng and Yip}(2007)}]{cheng2007trapped}
\bibinfo{author}{\bibfnamefont{C.}~\bibnamefont{Cheng}} \bibnamefont{and}
  \bibinfo{author}{\bibfnamefont{S.}~\bibnamefont{Yip}},
  \bibinfo{journal}{Phys. Rev. B} \textbf{\bibinfo{volume}{75}},
  \bibinfo{pages}{14526} (\bibinfo{year}{2007}).

\bibitem[{\citenamefont{Shin}(2008)}]{shin2008determination}
\bibinfo{author}{\bibfnamefont{Y.}~\bibnamefont{Shin}}, \bibinfo{journal}{Phys.
  Rev. A} \textbf{\bibinfo{volume}{77}}, \bibinfo{pages}{41603}
  (\bibinfo{year}{2008}).

\bibitem[{\citenamefont{Ho and Zhou}(2009)}]{ho2009obtaining}
\bibinfo{author}{\bibfnamefont{T.}~\bibnamefont{Ho}} \bibnamefont{and}
  \bibinfo{author}{\bibfnamefont{Q.}~\bibnamefont{Zhou}},
  \bibinfo{journal}{Nat. Phys.} \textbf{\bibinfo{volume}{6}},
  \bibinfo{pages}{131} (\bibinfo{year}{2009}).

\bibitem[{\citenamefont{Nascimbene et~al.}(2010)\citenamefont{Nascimbene,
  Navon, Jiang, Chevy, and Salomon}}]{nascimb2010exploring}
\bibinfo{author}{\bibfnamefont{S.}~\bibnamefont{Nascimbene}},
  \bibinfo{author}{\bibfnamefont{N.}~\bibnamefont{Navon}},
  \bibinfo{author}{\bibfnamefont{K.}~\bibnamefont{Jiang}},
  \bibinfo{author}{\bibfnamefont{F.}~\bibnamefont{Chevy}}, \bibnamefont{and}
  \bibinfo{author}{\bibfnamefont{C.}~\bibnamefont{Salomon}},
  \bibinfo{journal}{Nature} \textbf{\bibinfo{volume}{463}},
  \bibinfo{pages}{1057} (\bibinfo{year}{2010}).

\bibitem[{\citenamefont{Navon et~al.}(2010)\citenamefont{Navon, Nascimbene,
  Chevy, and Salomon}}]{navon2010equation}
\bibinfo{author}{\bibfnamefont{N.}~\bibnamefont{Navon}},
  \bibinfo{author}{\bibfnamefont{S.}~\bibnamefont{Nascimbene}},
  \bibinfo{author}{\bibfnamefont{F.}~\bibnamefont{Chevy}}, \bibnamefont{and}
  \bibinfo{author}{\bibfnamefont{C.}~\bibnamefont{Salomon}},
  \bibinfo{journal}{Science} \textbf{\bibinfo{volume}{328}},
  \bibinfo{pages}{729} (\bibinfo{year}{2010}).

\bibitem[{\citenamefont{Gross et~al.}(2010)\citenamefont{Gross, Shotan,
  Kokkelmans, and Khaykovich}}]{gross2010nuclear}
\bibinfo{author}{\bibfnamefont{N.}~\bibnamefont{Gross}},
  \bibinfo{author}{\bibfnamefont{Z.}~\bibnamefont{Shotan}},
  \bibinfo{author}{\bibfnamefont{S.}~\bibnamefont{Kokkelmans}},
  \bibnamefont{and}
  \bibinfo{author}{\bibfnamefont{L.}~\bibnamefont{Khaykovich}},
  \bibinfo{journal}{Phys. Rev. Lett.} \textbf{\bibinfo{volume}{105}},
  \bibinfo{pages}{103203} (\bibinfo{year}{2010}).

\bibitem[{\citenamefont{Zirbel et~al.}(2008)\citenamefont{Zirbel, Ni,
  Ospelkaus, Nicholson, Olsen, Julienne, Wieman, Ye, and
  Jin}}]{zirbel2008heteronuclear}
\bibinfo{author}{\bibfnamefont{J.}~\bibnamefont{Zirbel}},
  \bibinfo{author}{\bibfnamefont{K.}~\bibnamefont{Ni}},
  \bibinfo{author}{\bibfnamefont{S.}~\bibnamefont{Ospelkaus}},
  \bibinfo{author}{\bibfnamefont{T.}~\bibnamefont{Nicholson}},
  \bibinfo{author}{\bibfnamefont{M.}~\bibnamefont{Olsen}},
  \bibinfo{author}{\bibfnamefont{P.}~\bibnamefont{Julienne}},
  \bibinfo{author}{\bibfnamefont{C.}~\bibnamefont{Wieman}},
  \bibinfo{author}{\bibfnamefont{J.}~\bibnamefont{Ye}}, \bibnamefont{and}
  \bibinfo{author}{\bibfnamefont{D.}~\bibnamefont{Jin}},
  \bibinfo{journal}{Phys. Rev. A} \textbf{\bibinfo{volume}{78}},
  \bibinfo{pages}{13416} (\bibinfo{year}{2008}).

\bibitem[{\citenamefont{Giorgini et~al.}(1999)\citenamefont{Giorgini, Boronat,
  and Casulleras}}]{giorgini1999ground}
\bibinfo{author}{\bibfnamefont{S.}~\bibnamefont{Giorgini}},
  \bibinfo{author}{\bibfnamefont{J.}~\bibnamefont{Boronat}}, \bibnamefont{and}
  \bibinfo{author}{\bibfnamefont{J.}~\bibnamefont{Casulleras}},
  \bibinfo{journal}{Phys. Rev. A} \textbf{\bibinfo{volume}{60}},
  \bibinfo{pages}{5129} (\bibinfo{year}{1999}).

\bibitem[{\citenamefont{Leyronas and Combescot}(2007)}]{leyronas2007superfluid}
\bibinfo{author}{\bibfnamefont{X.}~\bibnamefont{Leyronas}} \bibnamefont{and}
  \bibinfo{author}{\bibfnamefont{R.}~\bibnamefont{Combescot}},
  \bibinfo{journal}{Phys. Rev. Lett.} \textbf{\bibinfo{volume}{99}},
  \bibinfo{pages}{170402} (\bibinfo{year}{2007}).

\bibitem[{\citenamefont{Krauth}(1996)}]{krauth1996quantum}
\bibinfo{author}{\bibfnamefont{W.}~\bibnamefont{Krauth}},
  \bibinfo{journal}{Phys. Rev. Lett.} \textbf{\bibinfo{volume}{77}},
  \bibinfo{pages}{3695} (\bibinfo{year}{1996}).

\bibitem[{\citenamefont{Castin and Dum}(1996)}]{castin1996bose}
\bibinfo{author}{\bibfnamefont{Y.}~\bibnamefont{Castin}} \bibnamefont{and}
  \bibinfo{author}{\bibfnamefont{R.}~\bibnamefont{Dum}},
  \bibinfo{journal}{Phys. Rev. Lett.} \textbf{\bibinfo{volume}{77}},
  \bibinfo{pages}{5315} (\bibinfo{year}{1996}).

\bibitem[{\citenamefont{Kagan et~al.}(1997)\citenamefont{Kagan, Surkov, and
  Shlyapnikov}}]{kagan1997evolution}
\bibinfo{author}{\bibfnamefont{Y.}~\bibnamefont{Kagan}},
  \bibinfo{author}{\bibfnamefont{E.}~\bibnamefont{Surkov}}, \bibnamefont{and}
  \bibinfo{author}{\bibfnamefont{G.}~\bibnamefont{Shlyapnikov}},
  \bibinfo{journal}{Phys. Rev. Lett.} \textbf{\bibinfo{volume}{79}},
  \bibinfo{pages}{2604} (\bibinfo{year}{1997}).

\bibitem[{inp()}]{inprep}
\bibinfo{note}{The details of the time-dependent analysis will be published
  elsewhere.}

\bibitem[{\citenamefont{Cowell et~al.}(2002)\citenamefont{Cowell, Heiselberg,
  Mazets, Morales, Pandharipande, and Pethick}}]{cowell2002cold}
\bibinfo{author}{\bibfnamefont{S.}~\bibnamefont{Cowell}},
  \bibinfo{author}{\bibfnamefont{H.}~\bibnamefont{Heiselberg}},
  \bibinfo{author}{\bibfnamefont{I.}~\bibnamefont{Mazets}},
  \bibinfo{author}{\bibfnamefont{J.}~\bibnamefont{Morales}},
  \bibinfo{author}{\bibfnamefont{V.}~\bibnamefont{Pandharipande}},
  \bibnamefont{and} \bibinfo{author}{\bibfnamefont{C.}~\bibnamefont{Pethick}},
  \bibinfo{journal}{Phys. Rev. Lett.} \textbf{\bibinfo{volume}{88}},
  \bibinfo{pages}{210403} (\bibinfo{year}{2002}).

\bibitem[{\citenamefont{Song and Zhou}(2009)}]{song2009ground}
\bibinfo{author}{\bibfnamefont{J.}~\bibnamefont{Song}} \bibnamefont{and}
  \bibinfo{author}{\bibfnamefont{F.}~\bibnamefont{Zhou}},
  \bibinfo{journal}{Phys. Rev. Lett.} \textbf{\bibinfo{volume}{103}},
  \bibinfo{pages}{25302} (\bibinfo{year}{2009}).

\bibitem[{\citenamefont{Lee and Lee}(2010)}]{lee2010universality}
\bibinfo{author}{\bibfnamefont{Y.-L.} \bibnamefont{Lee}} \bibnamefont{and}
  \bibinfo{author}{\bibfnamefont{Y.-W.} \bibnamefont{Lee}},
  \bibinfo{journal}{Phys. Rev. A} \textbf{\bibinfo{volume}{81}},
  \bibinfo{pages}{063613} (\bibinfo{year}{2010}).

\bibitem[{\citenamefont{Ho}(2004)}]{ho2004universal}
\bibinfo{author}{\bibfnamefont{T.}~\bibnamefont{Ho}}, \bibinfo{journal}{Phys.
  Rev. Lett.} \textbf{\bibinfo{volume}{92}}, \bibinfo{pages}{90402}
  (\bibinfo{year}{2004}).

\bibitem[{\citenamefont{Inguscio et~al.}(2006)\citenamefont{Inguscio, Ketterle,
  and Salomon}}]{inguscio2006ultracold}
\bibinfo{author}{\bibfnamefont{M.}~\bibnamefont{Inguscio}},
  \bibinfo{author}{\bibfnamefont{W.}~\bibnamefont{Ketterle}}, \bibnamefont{and}
  \bibinfo{author}{\bibfnamefont{C.}~\bibnamefont{Salomon}},
  \bibinfo{journal}{Proceedings of the International School of Physics "Enrico
  Fermi", Course CLXIV, Varenna}  (\bibinfo{year}{2006}).

\bibitem[{uni()}]{unitaryradiusscaling}
\bibinfo{note}{At unitarity, the cloud radius $R$ would scale as $N^{1/6}\xi^{1/4}$.
  The normalization $R^*$ used in Fig.4 scales as $N^{1/5}$ so that
  $R/R^*\propto\xi^{1/4}N^{-1/30}$ very slowly depends on atom number. In order to take into account the changing atom number near unitarity and obtain a safe experimental lower bound on $\xi\propto (R/R^*)^4 N^{2/15}$, we minimize both $R/R^*$ and $N^{2/15}$. This is done by taking for $R/R^*$ the initial atom number (empty circles in the inset of Fig.4), and the final, for $N^{2/15}$.}

\end{thebibliography}

\end{document}